\documentclass[aps,floats,floatfix,showpacs,twoside,preprint,superscriptaddress]{revtex4}

\usepackage{latexsym}
\usepackage{amsfonts}
\usepackage{graphicx}
\usepackage{amsbsy}
\usepackage{float}
\usepackage{amsmath}

\begin{document}
\vspace{-20mm}

\title{ Multiple Tipping Points and Optimal Repairing in Interacting Networks }

\author{Antonio~Majdandzic}
\affiliation{Center for Polymer Studies and Department of Physics, Boston
University, Boston, MA 02215, USA}

\author{Lidia A. Braunstein} 
\affiliation{Instituto de Investigaciones
  F\'isicas de Mar del Plata (IFIMAR), Physics Department, Universidad
  Nacional de Mar del Plata-CONICET, Funes 3350, (7600) Mar del Plata,
  Argentina}
\affiliation{Center for
  Polymer Studies and Department of Physics, Boston University,
  Boston, MA 02215, USA}

\author{Chester Curme}
\affiliation{Center for Polymer Studies and Department of Physics, Boston
University, Boston, MA 02215, USA}

\author{Irena Vodenska}
\affiliation{Center for Polymer Studies and Department of Physics, Boston
University, Boston, MA 02215, USA} 
\affiliation{Administrative Sciences Department, Metropolitan College,
  Boston University, Boston, Massachusetts 02215 USA}

\author{Sary Levy-Carciente}
\affiliation{Center for Polymer Studies and Department of Physics, Boston
University, Boston, MA 02215, USA}
\affiliation{Economics and Social Sciences Faculty, Central University
  of Venezuela, Caracas, Venezuela} 

\author{H.~Eugene~Stanley}
\affiliation{Center for Polymer Studies and Department of Physics, Boston
University, Boston, MA 02215, USA}

\author{Shlomo Havlin}
\affiliation{Center for Polymer Studies and Department of Physics, Boston
University, Boston, MA 02215, USA}
\affiliation{
Department of Physics, Bar-Ilan University,
 52900 Ramat-Gan, Israel}

\date{Version February 8, 2015.}

\begin{abstract} 

{\bf Systems that comprise many interacting dynamical networks, such as
  the human body with its biological networks or the global economic
  network consisting of regional clusters, often exhibit complicated
  collective dynamics.  To understand the collective behavior of such
  systems, we investigate a model of interacting networks exhibiting the
  fundamental processes of failure, damage spread, and recovery. We find
  a very rich phase diagram that becomes exponentially more complex as the
  number of networks is increased. In the simplest example of $n=2$
  interacting networks we find two critical points, 4 triple points, 10
  allowed transitions, and two ``forbidden'' transitions, as well as
  complex hysteresis loops.
  Remarkably, we find that triple points play the dominant role in 
  constructing the optimal repairing strategy in damaged interacting 
  systems.  To support our model, we analyze an example of
  real interacting financial networks and find evidence of rapid
  dynamical transitions between well-defined states, in agreement with
  the predictions of our model.}

\end{abstract}

%\pacs{??}

\maketitle

Most real networks are not isolated structures but interact with other
network structures. As a result, much research has been focused recently
on the dynamics of interdependent \cite{BookShlomo, Sergey, Gao, ArenasNP, Reis, Donges, Vespi1, Bashan} and multilayer \cite{Bian, Glee, Bocca} networks.  Recent
studies on network repair \cite{Derek, Liu, BP2} have shown the
importance of recovery of nodes as a process which leads to reverse transitions,
hysteresis effects, and such phenomena as spontaneous recovery
\cite{Derek, BP}.

The cardiovascular and nervous systems in the human body are examples of
two dynamically interacting physiological networks \cite{Bashan2}. Diseases often
result from complex pathological conditions that involve a dynamical
interaction with positive or negative feedback between different
functional subsystems in the body. Similarly, in the global economy
there is a hierarchy of clustered and more tightly connected countries,
often grouped geographically, that are further interconnected to one
large global interacting economic and financial network \cite{Guido, Chester, Vodenska}.  To understand
the behavior of these systems using network science, we develop a model
of interacting networks with nodes that can recover from failure and we
examine the resulting phase diagram. The phase diagram we find is very rich
and contains a number of tipping points (critical points \cite{Dor, Newman1, CohenHavlin, Barabasi2002}, 
triple points \cite{Lucas, Lucas2014, Nunes} and transition lines).
The number of critical points grows linearly as the number of interacting networks in the system is increased, 
while the number of triple points and transition lines grows exponentially.  We present our
method and the results in detail for the simplest case of $n=2$
interacting networks, which can be easily generalized to any number of
interacting networks.

Our model of a generic system consisting of interacting dynamical
networks captures the important events found in real-world interacting
networks, i.e., node failure \cite{Albert, Schneider, Parshani, HelbingPRL},
systemic damage propagation\cite{Helbing}, and node recovery
\cite{Derek, BP, Kertesz}. In our model we first describe the
structure of the system and then describe the rules governing the
dynamic behavior of the processes occurring within the system.

The structure of our system for the $n=2$ case is modeled as follows. 
We start with two isolated networks, network A
and network B, and for simplicity we assume that both networks have the same
number of nodes $N$ and the same degree distribution $f(k)$ (these
assumptions can be relaxed with a cost of additional complication, but
the results are qualitatively similar).  We assume that within each
network the nodes are randomly connected. Now, to allow networks A and B to
interact, we introduce interdependency links that connect nodes across
the two networks \cite{Sergey}. This can be achieved in different
ways, and we use a simple one-to-one dependency: each node in network A is
dependent on exactly one node in network B, and vice versa. The pairs of nodes
of both networks are chosen randomly.

The dynamic behavior of our system is governed by two categories of
event---failure and recovery---and we assume that every node is in
either a failed or an active state. Node failure can result from
internal failure or from the spread of damage from neighbor nodes in
either the same network or the interdependent network.  We thus assume
that there are three ways a node can fail: (i) internally induced
failure, when a node's internal integrity has been compromised, e.g., an
organ in the body can fail due to a malfunction within the organ or a company
can fail due to bad management, (ii) externally induced failure through
failure propagation due to connections with failed nodes within the
node's own network, and (iii) failure induced through the dependency 
link as a result of being dependent on a failed node from another (opposite)
network. Apart of these three types of failures, we assume the existence 
of associated simple recovery processes for every type of failure. 
We specify quantitatively each of these processes below.

\begin{itemize}

\item[{(i)}] {\bf Internal failure} (I). We assume that in both networks
  any node can fail due to internal problems, independent of other
  nodes. For each node in network A we assume that there is probability
  $p_A~dt$ that the node will fail internally during any time period
  $dt$.  The equivalent parameter in network B is $p_B$.

\item[{(ii)}] {\bf External failure} (E). Every node in network A and
  network B is connected by links to nearby nodes in its own
  network. These nodes constitute the node's neighborhood. The number of
  links a node has within the network indicates its {\it degree\/} or
  {\it connectivity}, denoted by $k$.  If a large number of nodes in a
  node's neighborhood have failed, i.e., if the neighborhood is
  substantially damaged, we assume that the probability that the node
  itself will fail is increased. As in Refs.~\cite{Derek} and
  \cite{Watts2}, we use a threshold rule to define a {\it substantially
    damaged\/} neighborhood, which is a neighborhood containing $\le m$
  active nodes, where $m$ is a fixed integer threshold. If node $j$ has
  $> m$ active neighbors during time $dt$, we consider its neighborhood
  to be ``healthy'' and there is no risk of external failure. On the
  other hand, if $j$ has $\le m$ active neighbors during time $dt$,
  there is a probability $r_A~dt$ (for network A) or $r_B~dt$ (for
  network B) that node $j$ will externally fail.  (For an explanation of
  why $r_A$ and $r_B$ are not set to 1 and why they are necessary, see
  Note 1 in Methods).
   
\item[{(iii)}] {\bf Dependency failure} (D). In the case of two
  interdependent networks (A and B) we assume that each node in the
  first network is dependent on a node in the second network via an
  interdependent link, and vice versa. We assume that if one node in the
  pair fails there is a finite (but not 100\%) probability, $r_d~dt$,
  that during time $dt$ the other node in the pair will fail as well.
  This represents the probability that the damage will spread through the
  interdependency link.

\item[{(iv)}] {\bf Recovery}. We assume that there is a reversal
  process, a recovery from each of these three types of failure. A node
  recovers from an {\it internal\/} failure after a time period $\tau
  \ne 0$, it recovers from an {\it external\/} failure after time
  $\tau'$, and from a {\it dependency\/} failure after time $\tau''$.
  In simulations, and without loss of generality, we use $\tau=100$, and
  for simplicity we set $\tau'=\tau''=1$ to take into account the
  assumption that real-world systems usually require a longer time
  period to recover from internal problems (physical faults) then from a
  lack of environmental support.  Changing the numerical values however,
  does not introduce any qualitative difference.
 
 \item[{(iv)}] {\bf Activity notation.} Every node is in one of two
   states: active or failed. A node is considered active in the observed moment,
   if it is not experiencing internal (I), external (E), or dependency (D)
   failure.
   
\end{itemize}

\section{Results}

\subsection{Mean field theory}

We characterise this system by studying the order parameters chosen naturally
as the fraction of active nodes in network A and network B, $z_A$ and
$z_B$, respectively. For purposes of simplifying the calculation, however, we first
concentrate on the complementary and equally intuitive fraction of {\it
  failed\/} nodes $a_A$ and $a_B$, in networks A and B respectively
($a_{A}=1-z_{A}$, $a_{B}=1-z_{B}$).

Using the mean field theory presented in Methods, Note 2, we
obtain two coupled equations that connect $a_A$ and $a_B$, which the
system must satisfy in the equilibrium
\begin{eqnarray}
a_A&=&p_A^*+r_d a_B(1-p_A^*) + \sum_{k}f(k) F(k, a_{A})[r_A-p_A^* r_A
  -r_A r_d a_B+p_A^*r_A r_d a_B]\label{e1}\\ 
a_B&=&p_B^*+r_d a_A(1 -p_B^*) + \sum_{k}f(k) F(k, a_{B})[r_B-p_B^* r_B
  -r_B r_d a_A+p_B^*r_B r_d a_A] \label{e2} 
\end{eqnarray}

Here $F(k,x)=\sum_{j=0}^{m}{{k\choose{j}} x^{k-j}{(1-x)}^j}$, and we
have also introduced simplifying parameters $p_A^*\equiv e^{-p_A \tau}$
and $p_B^*\equiv e^{-p_B \tau}$ to make the equations more elegant and to
reduce the number of parameters by replacing $p_A$, $p_B$, and $\tau$
that appear as a product. We find that the parameters $p_A^*$ and
$p_B^*$ are very convenient to work with because they correspond to the fraction 
of internally failed nodes in network A and network B, respectively.

Despite the seeming complexity of Eqs.~(1) and (2), note that there are
only two unknown variables, $a_A$ and $a_B$, and that all other parameters are
fixed. These two equations define two curves
in the ($a_A, a_B$) plane.

\begin{figure} [!htbp]
%\centering \includegraphics[width=0.30\textwidth]{venn.jpg}
\centering \includegraphics[width=0.45\textwidth]{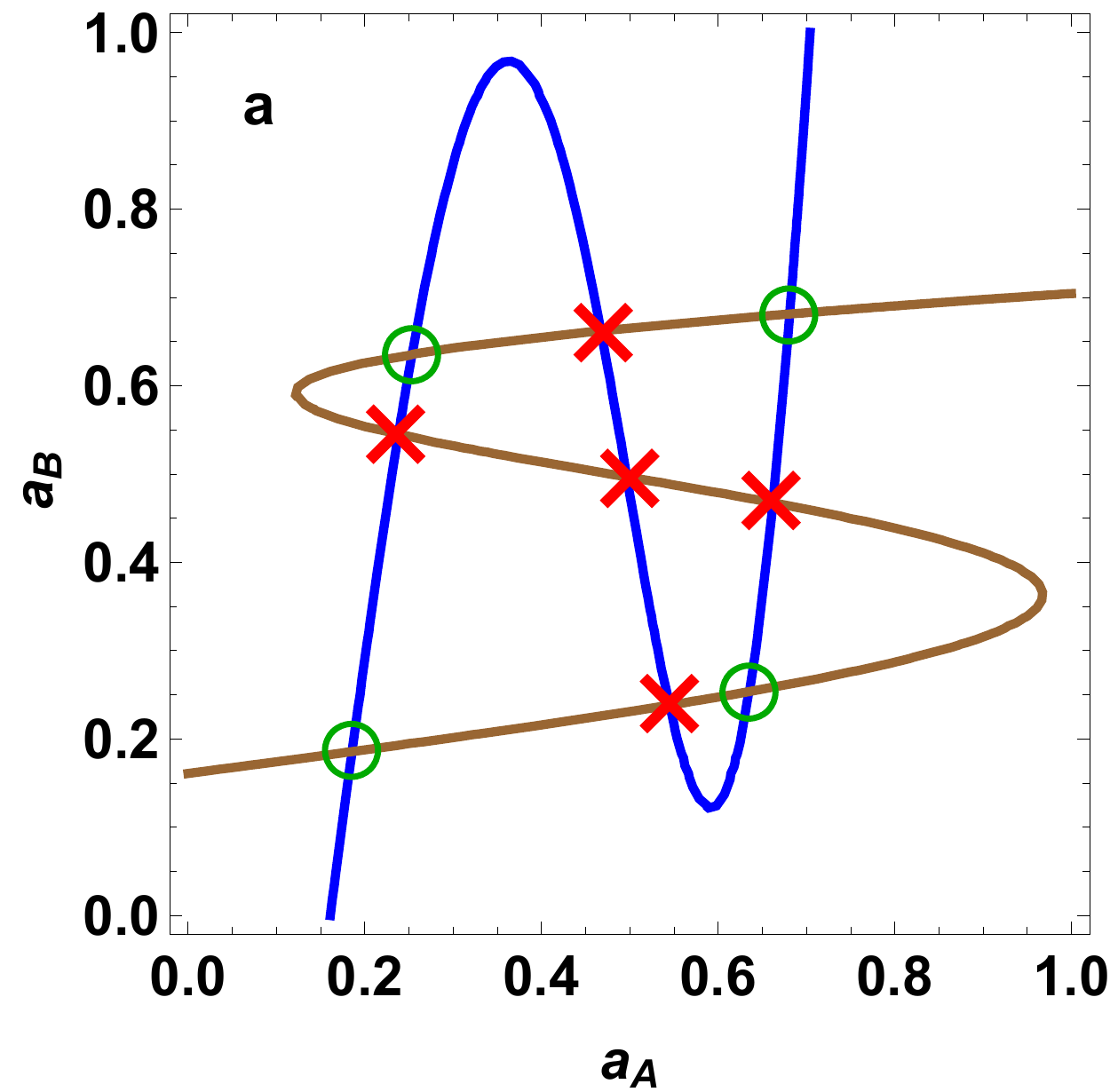}
\centering \includegraphics[width=0.45\textwidth]{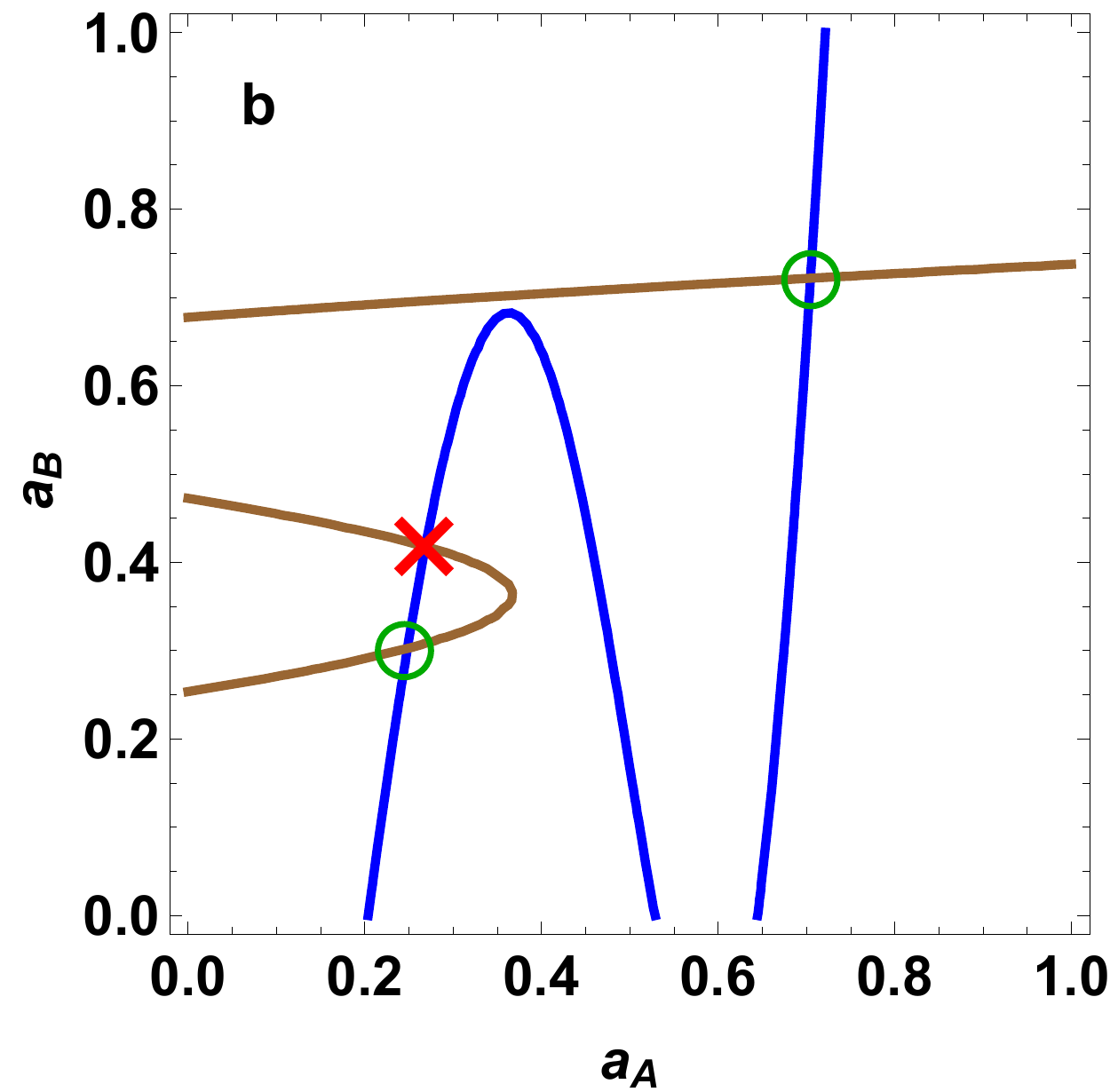}
\caption{ {\bf Graphical representations of the mean field equations for
    a system with two interdependent networks ($k=16$, $m=8$).}  {\bf a)
  } The blue and brown curves represent Eq. (1) and Eq. (2), respectively,
  for $p_A^*=p_B^*=0.16$, $r_A=r_B=0.60$ and $r_d=0.15$. There are nine
  intersections, representing mathematical solutions for network
  activities $a_A$ and $a_B$. Four of them are stable solutions (green
  circles) representing physical states that we also observe in our
  simulations, and five are unstable solutions (red crosses).  {\bf
    b) } Example for $p_A^*=0.20$, $p_B^*=0.24$, $r_A=r_B=0.60$ and
  $r_d=0.15$. Here we obtain two stable and one unstable solutions. The two
  stable solutions correspond to 11 state (both networks are at high activity)
  and 22 state (both networks are at low activity).  }
   \label{1}
\end{figure}

Figure~1a shows a graphical representation of the curves for a random
regular \cite{CohenHavlin} network (in which all the nodes have the same
degree) with degree of $k=16$ and threshold $m=8$, for the symmetric
parameter values $p_A^*=p_B^*=0.16$, $r_A=r_B=0.60$, and
$r_d=0.15$. The size of each network is $N=2 \times 10^4$.  The blue
curve is a graphical representation of Eq.~(\ref{e1}), and the brown
curve is defined by Eq.~(\ref{e2}). The curves, like two ``ropes,''
create a ``knot'' that can have up to nine intersections, representing
mathematical solutions of the system of equations.  Not all of these
solutions represent observable physical states, however. 
Some of them turn out to be unstable and we need to discard them. 
To see that, observe one of the curves in Fig.~1a, for example the blue curve 
described by Eq.~(1).
We can think of this curve as describing the fraction of failed nodes $a_A$ 
in network A as a function of $a_B$ (the fraction of failed nodes in network B),
keeping everything else fixed. If we increase damage done to network B 
(i.e. we increase $a_B$) and keep everything else constant, 
some damage will undoubtedly spread to network A. Thus we expect that when
$a_B$ is increased, $a_A$ must also increase (it would be very unusual if one network
improves its activity as a result of damaging the other network, in our model where
activities of the two networks are positively correlated).
We conclude that the parts of the blue and brown curve that
produce physical solutions are only those where $a_A$ and $a_B$ increase
together or decrease together along the curve.
This elimination leaves only four states in Fig.~1a that are stable (green
circles), while the other five states are unstable (red crosses), for this
particular choice of parameters. Generally, for any choice of
parameters, we have between one and four physical
states. Figure~1b shows the scenario for the same network system when
$p_A^*=0.20$, $p_B^*=0.24$, $r_A=r_B=0.60$, and $r_d=0.15$. In this case
we have two stable states and one unstable.  This mean field theory
calculation agrees well with the states that we observe in our
simulations, as we will demonstrate below.

Note that our choice of $r_d$ value is quite limited. If $r_d$ is too large,
we find that the damage spreads through dependency links extremely efficiently
and the only possible stable state is total system collapse.  The extreme
vulnerability of interdependent networks is well-known
\cite{Sergey, Parshani}.
Because there is always at least one functional
stable state in biological or man-made systems, total system
collapse as the only stable state is not realistic. 
Thus we need the $r_d$ parameter to
''soften'' the dependency links \cite{Parshani} and allow a more realistic behavior.

The four physical solutions found above correspond to the following four scenarios:
(i) when there is high activity in both network A and network B
  (denoted ``11'' or ``up-up''),
(ii) when there is high activity in network A and low activity
  in network B (``12'' or ``up-down''),
(iii) when there is low activity in network A and high activity
  in network B (``21'' or ``down-up''), and
(iv) when there is low activity in both network A and network B 
(``22'' or ``down-down'').

Depending on the parameters, we obtain between one and four stable
states. Each of the states exists in a certain volume of the
multi-dimensional space of parameters.  Results of the mean field theory
calculation for a particular set of parameters are presented in
Fig.~2a-d as a phase diagram with four layers.  Figure~2 shows the
regions in which each of the four states exist in the ($p_A^*$, $p_B^*$)
parametric sub-space, when other parameters are fixed at values
$r_A=r_B=0.60$ and $r_d=0.15$, with $k$ and $m$ remaining the same as before.  

For example, in Fig.~2a the green area indicates the region where the 11 state exists. 
This state (phase) is bounded with a smooth boundary of three colors. If the
boundary is crossed (by increasing $p_A^*$ and $p_B^*$), the system
makes a transition to either state 12 (if the orange line is crossed),
state 22 (if the blue line is crossed), or state 21 (if the purple line
is crossed). The arrows indicate transitions. In Fig.~2a there are
two triple points (black points) that mark the change in the
transition type and where three different states can exist. The
blue area in Fig.~2b indicates the 22 state. This layer of
the phase diagram has two triple points as well, and three possible
transitions ($22\rightarrow12$, $22\rightarrow11$,  and $22\rightarrow21$).

Figures 2c and 2d show the regions of state 21 (purple) and state 12
(orange), respectively.  Each has two different transitions and one
 critical point.  For example, there are two possible ways out of state
21 (Fig.~2c): by a transition to the 11 (green arrow) state or the 22
(blue arrow) state.  Note that the different state regions (Figs.~2a,
2b, 2c, and 2d) are not disjoint sets but there is an overlap, resulting in 2-fold,
3-fold, or even 4-fold hysteresis regions.

The state in which the system is found depends on the initial
conditions or the system's past.  There are a total of 10
different transitions ($11\rightarrow12$, $11\rightarrow22$,
$11\rightarrow21$, $12\rightarrow11$, $12\rightarrow22$,
$21\rightarrow11$, $21\rightarrow22$, $22\rightarrow12$,
$22\rightarrow21$ and $22\rightarrow11$) that connect different layers
of the phase diagram (states 11, 12, 21, and 22), much like elevators
connecting different floors.  Transitions $12\rightarrow21$ and
$21\rightarrow12$ are the only missing (``forbidden'') combinations.
Although regions 12 and 21 do overlap, there is no a direct transition
connecting these two states.  These transitions would correspond to the
unusual combination in which one network recovers (transitions to a
higher activity) and simultaneously the other network fails. Thus a
transition from state 12 to state 21 requires the use of an intermediate
state (11 or 22). A more detailed discussion of the absence of these two
transitions can be found in Methods, part 3. The set of all allowed and
forbidden transitions is presented in Fig.~2e.  The total phase diagram
(all four layers on top of each other) is presented in Fig.~3.  Here,
color lines represent the boundaries of four states, with each color
corresponding to the boundary of one state, e.g., the green line is a
boundary of the 11 state. Note that there is a small central ``window''
where all four states are possible.

\begin{figure} [!htbp]
\centering \includegraphics[width=0.38\textwidth]{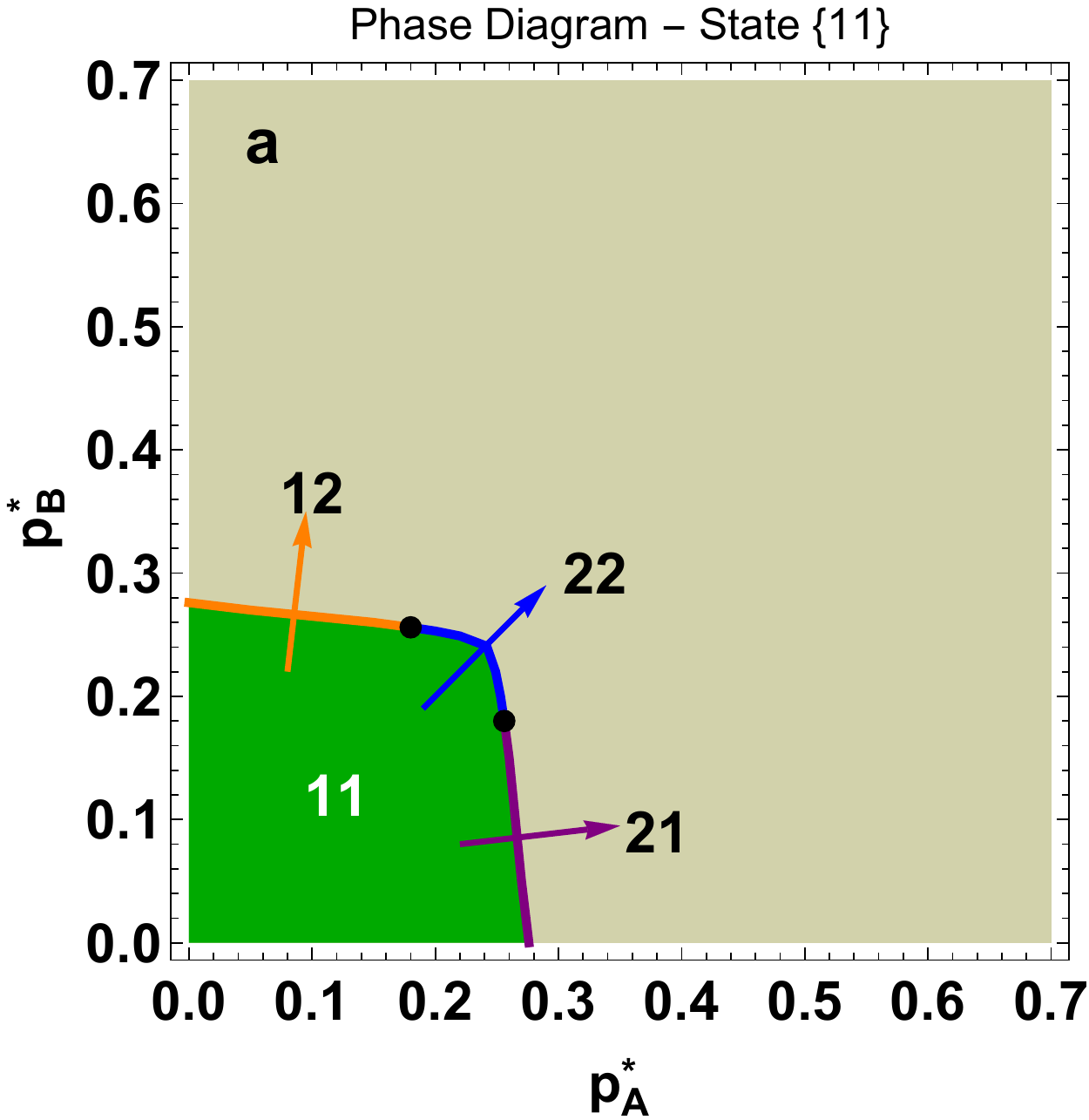}
\centering \includegraphics[width=0.38\textwidth]{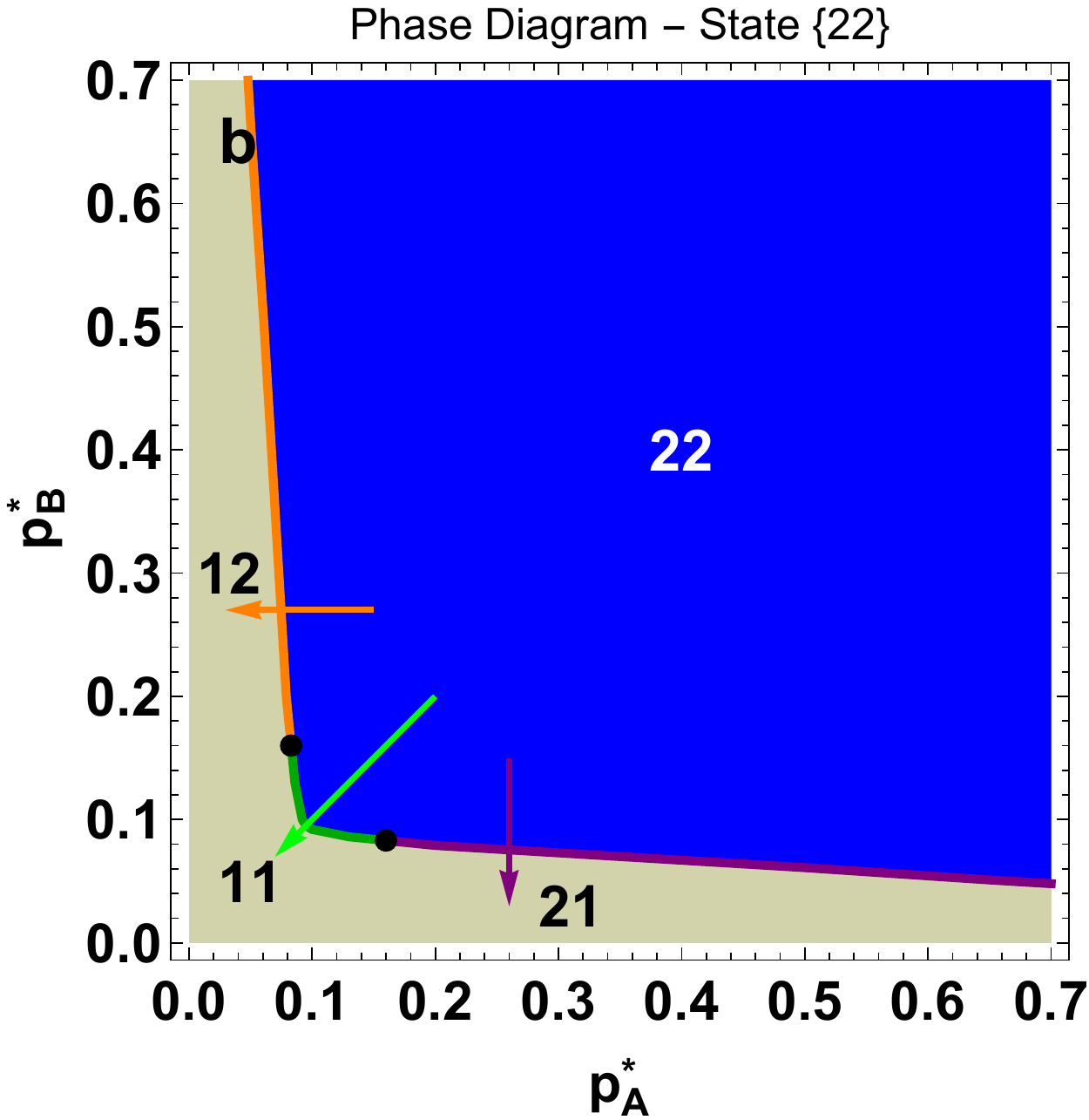}\\
\centering \includegraphics[width=0.38\textwidth]{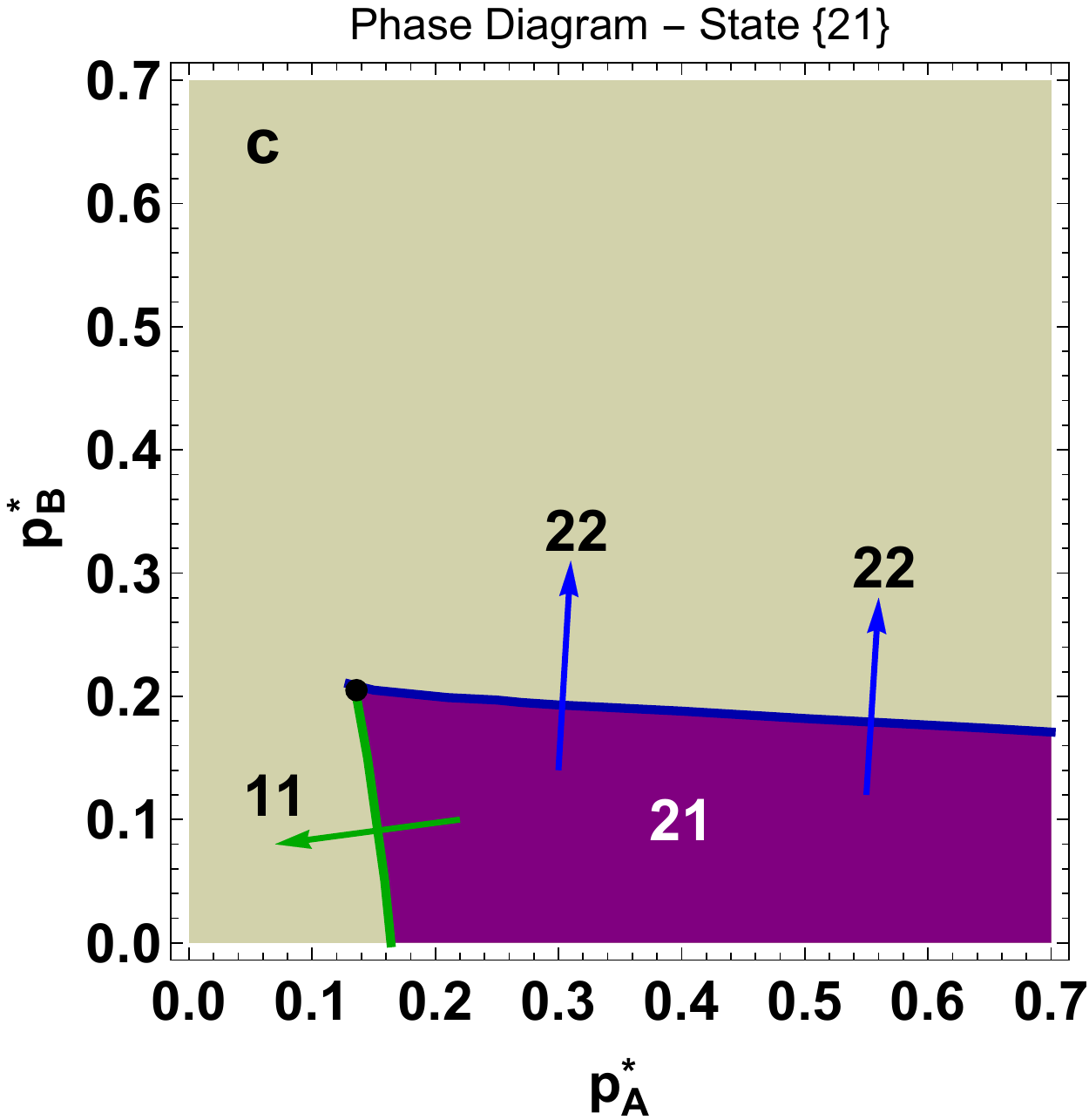}
\centering \includegraphics[width=0.38\textwidth]{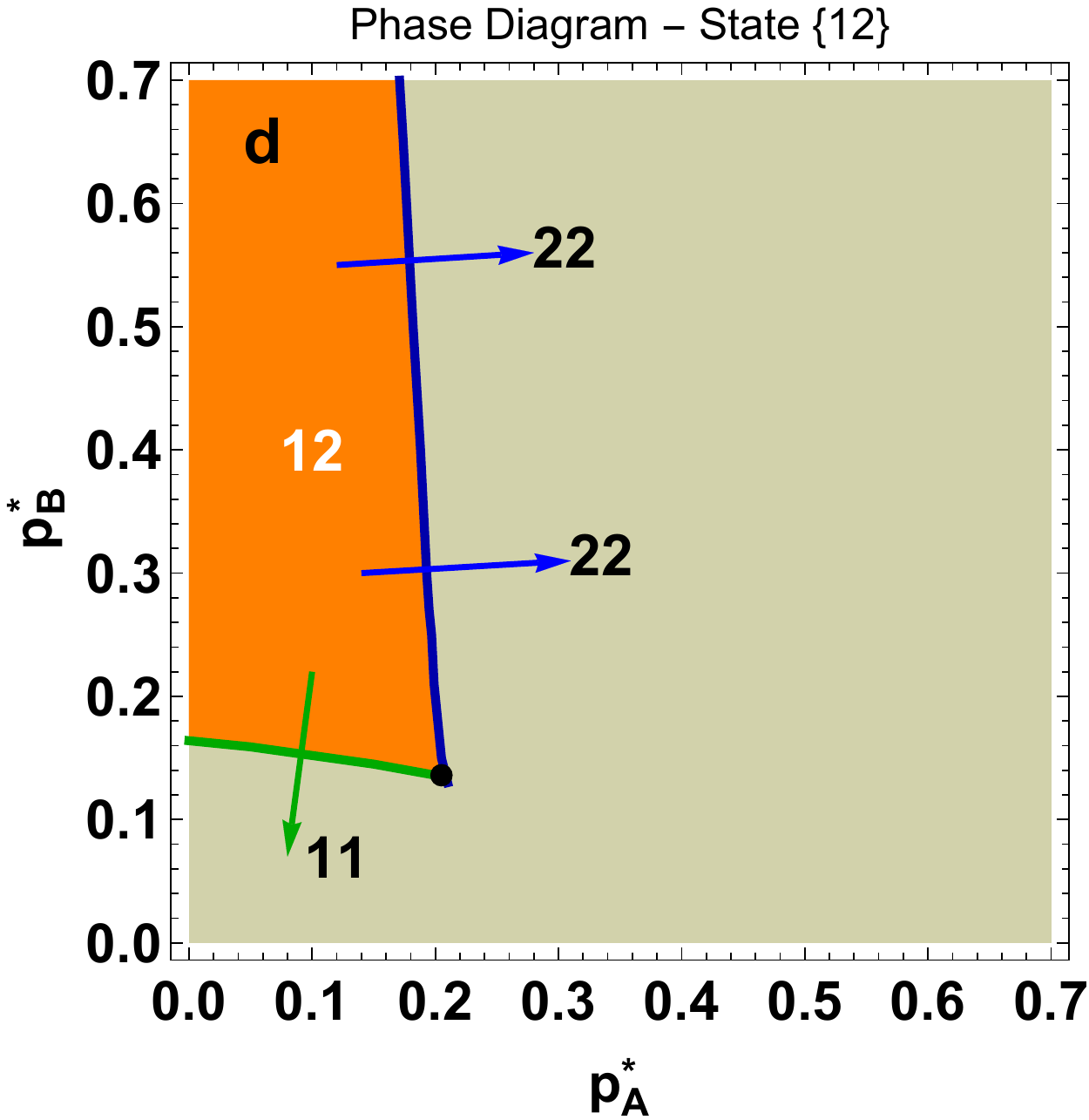}\\
\centering \includegraphics[width=0.23\textwidth]{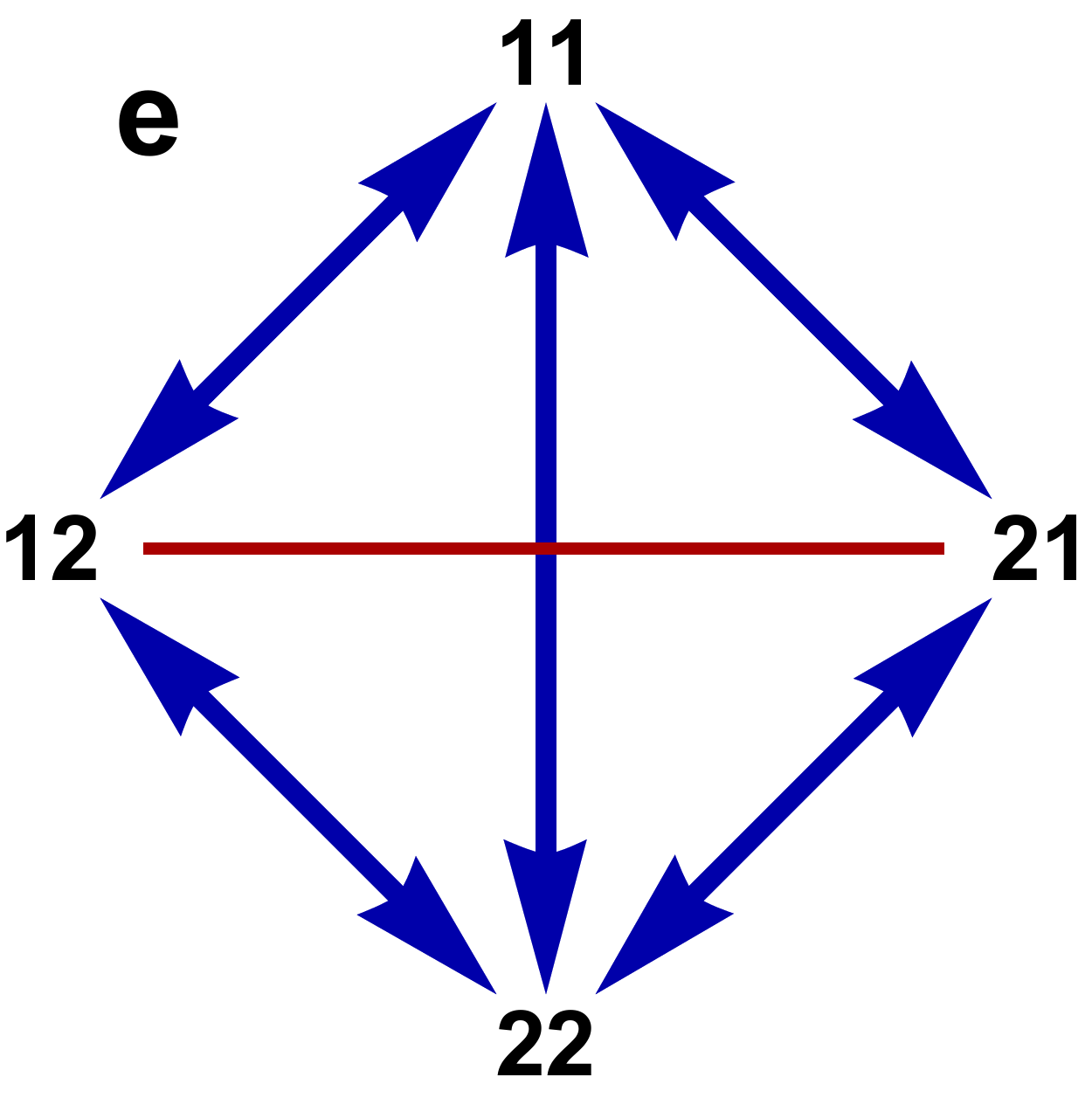}
\caption{ {\bf Four layers of the phase diagram and the transitions
    connecting them. }  {\bf a) } Region of 11 state, in green. Possible
  transitions are $11\rightarrow12$ (orange line), $11\rightarrow22$
  (blue line) and $11\rightarrow21$ (purple line). This layer of the
  phase diagram has two triple points, marked as black points. 
  {\bf b) } Region of 22 state (
  blue), with two triple points and three transitions.  {\bf c) } Region
  of 21 state (purple), with two transition lines (to 11 and 22 state)
  that merge in a critical point.  {\bf d) } Region of 12 state
  (orange), with two transition lines (to 11 and 22 state) that merge in
  a critical point.  {\bf e) } Illustration showing states (11, 12, 21 and
  22) with allowed (blue arrows) and "forbidden" (red line)
  transitions.  }
   \label{2}
\end{figure}

\begin{figure} [!htbp]
\centering \includegraphics[width=0.80\textwidth]{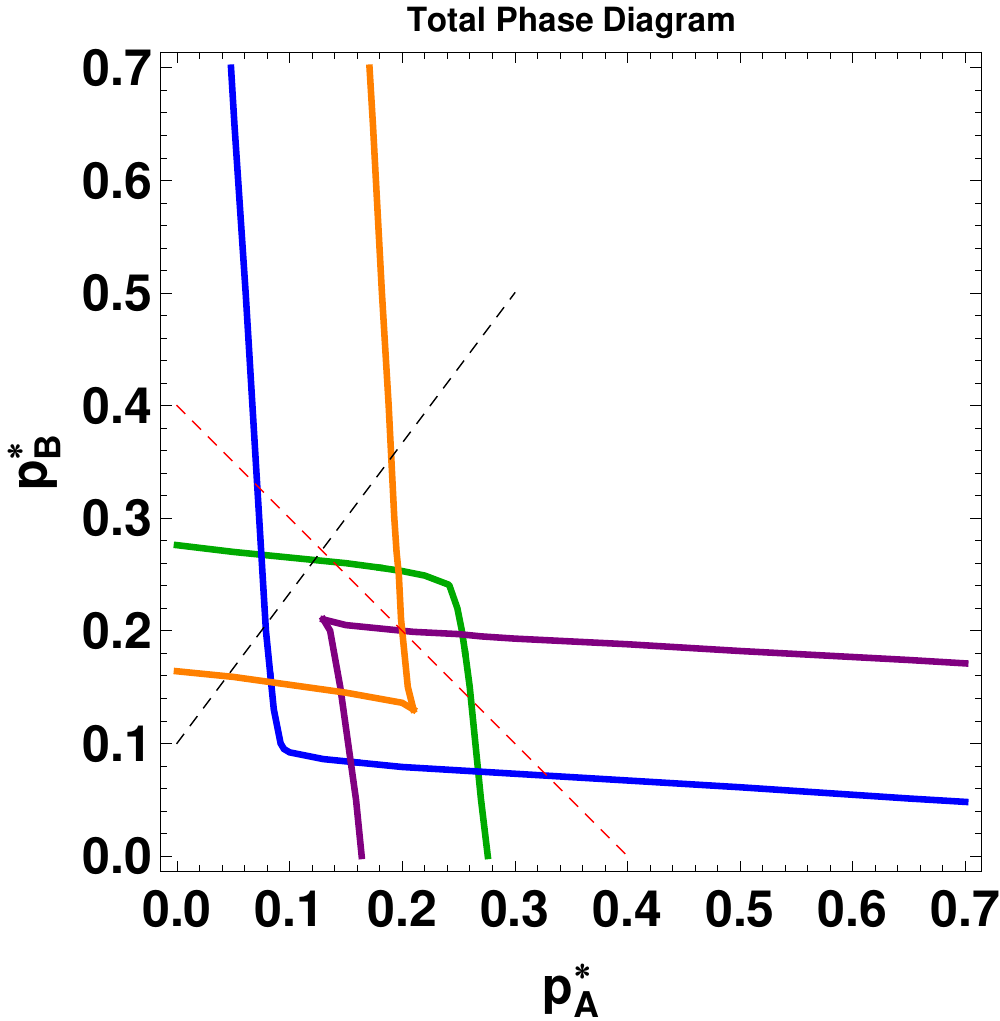}
\caption{ {\bf Total phase diagram, with all four layers.} Solid
  lines represent the border of region 11 (green), 22 (blue), 12
  (orange) and 21 (purple). Dashed lines represent cross-sections where
  we calculate the activity profile, shown in Figure 4.  }
   \label{3}
\end{figure}

We next can examine the activity profile for various cross-sections in
the phase diagram. In Figure~3 we choose two representative cross sections
(dashed straight lines) to measure activity $z_A=1-a_A$ as
$p_A^*$ and $p_B^*$ change.  The black dashed line is defined by the
equation $p_B^*=0.1+4/3 p_A^*$ and the red dashed line by
$p_B^*=0.4-p_A^*$.  Figure 4a~shows the activity measured in simulations
of network A as we move along the black dashed line, changing both
$p_A^*$ and $p_B^*$ and preserving the relation $p_B^*=0.1+4/3
p_A^*$. We perform simulations for various initial conditions and find (Fig.~4a)
three different states denoted by green, orange and blue colors, which
we identify as 11, 12, and 22 states, respectively. We find four different
transitions: $11\rightarrow12$, $12\rightarrow22$, $12\rightarrow11$,
and $22\rightarrow12$. The solid lines show the mean field theory (MFT)
prediction [Eqs.~(1) and (2)] for the activity of network A. The good agreement shows
that the mean field theory correctly captures all the properties of the
system. We note that qualitative agreement between the MFT and the
simulations is better for higher values of $k$, because for higher $k$
the fluctuations are smaller, which improves the accuracy of the
MFT. Figure~4b shows the activity when moving along the red dashed
line. Here we obtain four states and six different transitions.

\begin{figure} [!htbp]
\centering \includegraphics[width=0.62\textwidth]{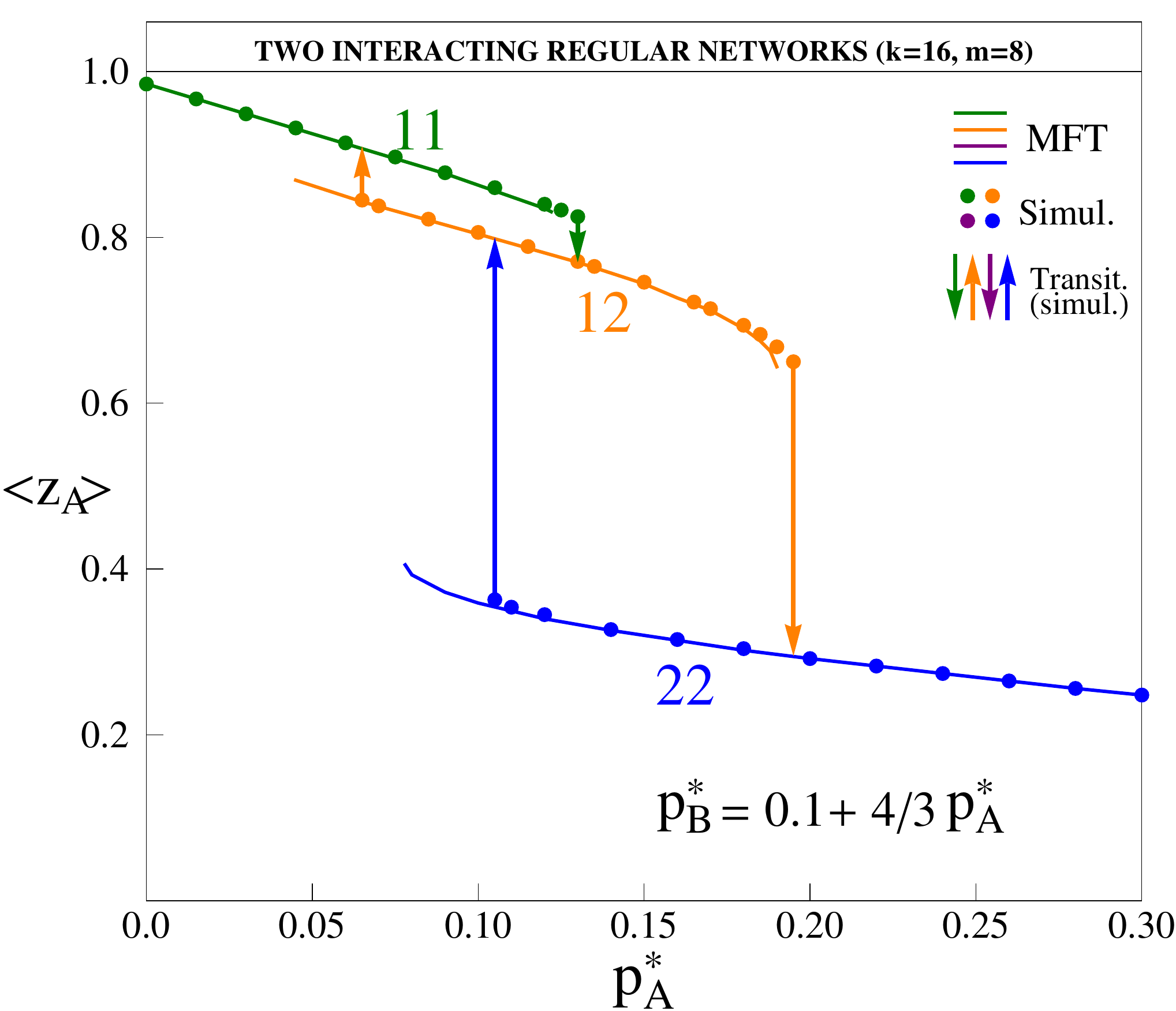}\\
\centering \includegraphics[width=0.62\textwidth]{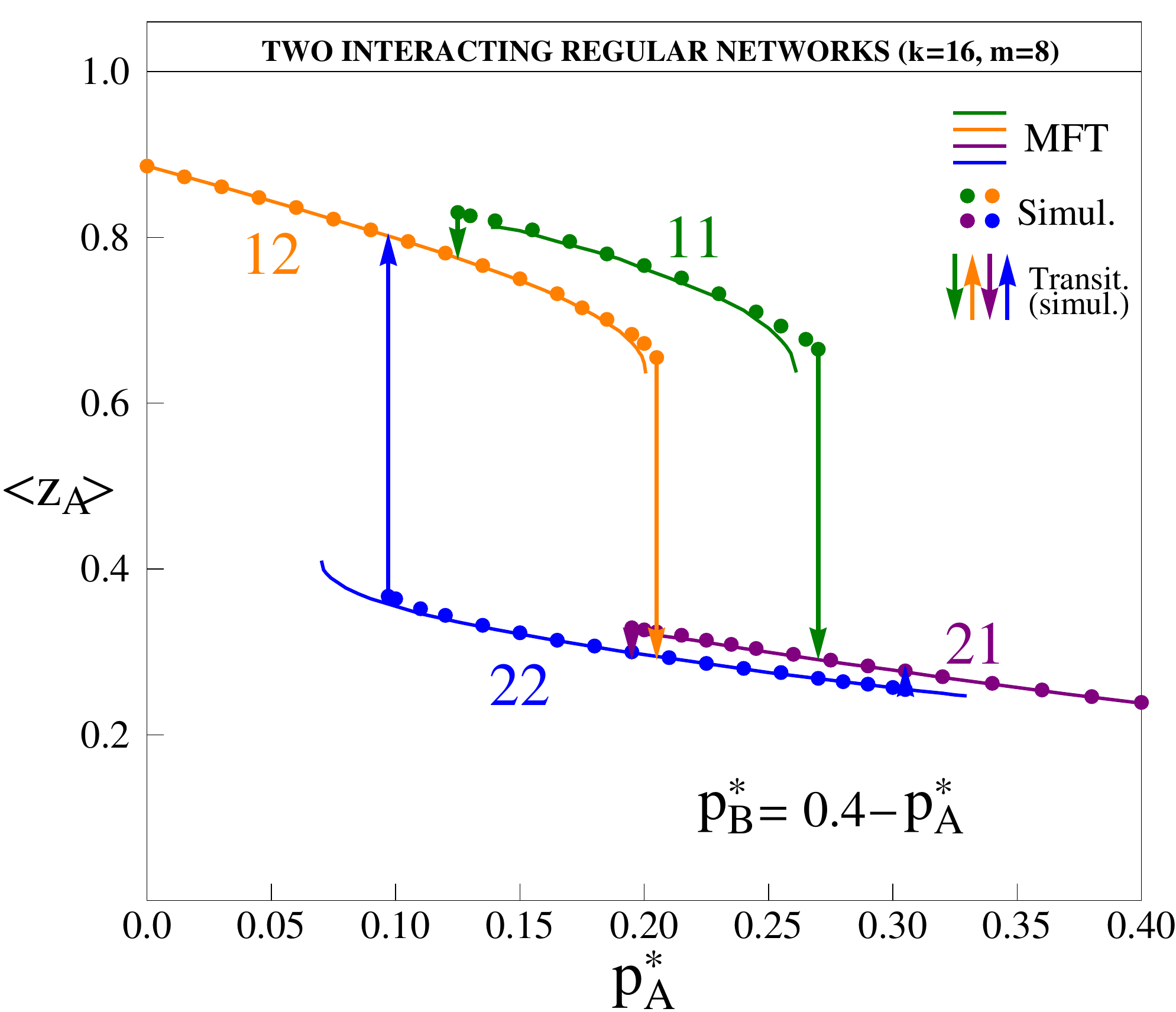} \\
\caption{ {\bf States, transitions and hysteresis loops for two activity
    profiles. } {\bf a) } Activity $z_A$ of network A, as measured in
  simulations (dots) and predicted by mean field theory (solid lines),
  along the cross section defined by the black dashed line in Fig
  3. Parameters $p_A^*$ and $p_B^*$ are changed, preserving the relation
  $p_B^*=0.1+4/3 p_A^*$. Transitions are denoted by arrows.  {\bf b) }
  Same for the cross section defined by $p_B^*=0.4-p_A^*$ (red dashed
  line in Fig. 3). Here we obtain 4 states and 6 different transitions,
  giving rise to more complex hysteresis loops.  }
   \label{4}
\end{figure}

The phase diagram of a system of $n=2$ interacting networks is much
richer than the phase diagram of a single network with damage and
recovery \cite{Derek}. 
The analytical results we presented here for $n=2$ can be
generalized to $n$ interacting networks in any topological
configuration, although as $n$ increases they become increasingly
difficult to visualize.
In general, a system with $n$ interacting
networks can have up to $2^n$ physical states. The maximum number of critical
points grows linearly with $n$ while the upper limit for the number of triple points grows
exponentially. 

\subsection{The problem of optimal repairing}

Knowing and understanding the phase diagram of interacting networks enable us to
answer some fundamental and practical questions.  A partially
or completely collapsed system of $n\geq2$ interacting networks in which
some of them are in the low activity state is a scenario common in
medicine, e.g., when diseases or traumas affect the human body and
a few organs are simultaneously damaged and need to be treated, and the
interaction between the organs is critical. It is also common in economics, 
when two or more coupled sectors of the economy \cite{Chester} experience simultaneous problems,
or when a few geographical clusters of countries experience economic difficulties.
The practical question that arises is: What is the most efficient strategy to
repair such a system? Many approaches are possible if resources are unlimited, 
but this is usually not the case and we would like to minimize the resources 
that we spend in the repairing process.

For simplicity, consider two interacting networks, both damaged (low activity).
Is repairing both networks simultaneously the more efficient approach, or
repairing them one after the other?  What is the minimum amount of
repair needed to make the system fully functional again?  In other
words, what is the minimum number of nodes we need to repair in order to
bring the system to the functional 11 (``up-up'') state, and how do we allocate
repairs between the two networks? An optimal repairing strategy is
essential when resources needed for repairing are limited or very
expensive, when the time to repair the system is limited, or when the
damage is still progressing through the system, threatening further
collapse, and a quick and efficient intervention is needed.

\begin{figure} [!htbp]
\centering \includegraphics[width=0.80\textwidth]{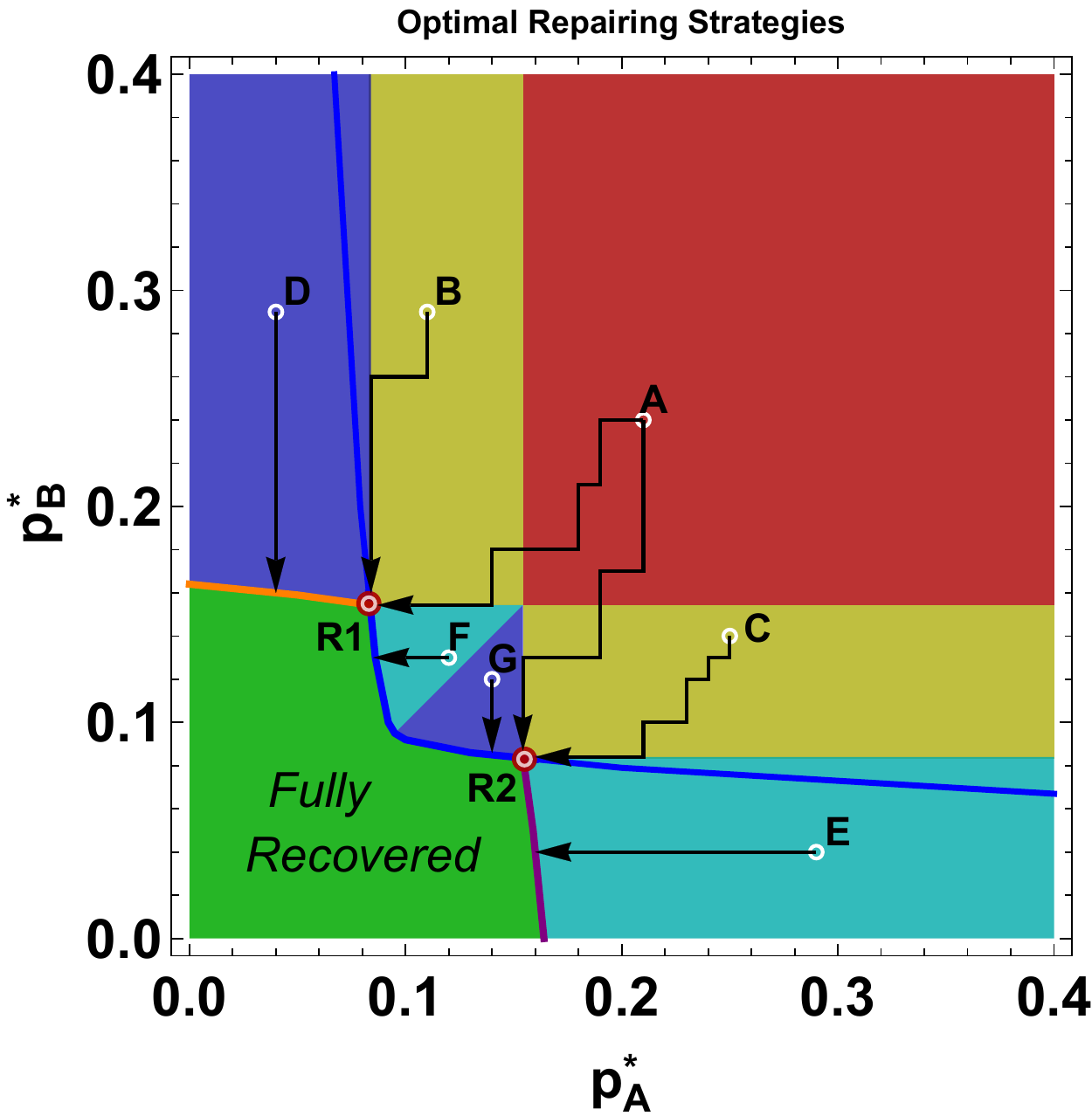}
\caption{ {\bf Optimal repairing strategies.} The solution to the
  problem of least expensive repairing corresponds to finding the
  minimal Manhattan distance from the point where the collapsed system
  is situated, and the border of the green region. In the red
  square region (point A for example), there are two solutions and it is
  equally optimal to reach any of the two triple points R1 and R2 by
  decreasing $p_A^*$ and $p_B^*$. In the yellow regions, it is optimal
  to reach only one triple point - R1 for the sector containing point B,
  or R2 for the sector containing point C. In the dark blue regions it
  is optimal to decrease $p_B^*$ only, and in the light blue regions it
  is optimal to decrease $p_A^*$ only. Note that triple points represent
  the solution of the optimal repairing for the warm color regions (red
  and yellow).  }
   \label{5}
\end{figure}

We show below that this problem is equivalent to finding the minimum
Manhattan distance between the point in the phase diagram where the
damaged system is currently situated, and the recovery transition lines
to the 11 region. The Manhattan distance between two points is defined
as the sum of absolute horizontal and vertical components of the vector
connecting the points, with defined vertical and horizontal
directions. It is a driving distance between two points in a rectangular
grid of streets and avenues. In our phase diagram, it is equal to
$|\Delta p_A^*|+|\Delta p_B^*|$. It turns out that two triple points of
the phase diagram play a very important role in this fundamental
problem. We find that these special points have a direct practical
meaning and are not just a topological or thermodynamic curiosity.

To show this, we start by making some simplifying but reasonable assumptions.
First, we assume that only internal failures can be
repaired by human hands, since these failures are physical faults in nodes (any
external and dependency failures and recoveries are ``environmental,''
and are a spontaneous recognition of the changing neighborhood of a
node).  We mentioned above that the parameters $p_A^*$ and $p_B^*$
correspond to fractions of internally failed nodes in networks A and B,
respectively. This implies that the number of internally failed nodes
repaired in, say, network A, is directly proportional to the change of
$p_A^*$. Hence repairing nodes in networks A and B means decreasing
$p_A^*$ or $p_B^*$. We also assume that these repairs are done fast
enough that there is only a small probability that the newly repaired
nodes will internally fail again before the repair process is completed.  The
total number of repaired nodes is therefore $N_{\rm rep}=N(|\Delta
p_A^*|+|\Delta p_B^*|)$, and it is proportional to the Manhattan distance between the
starting and final point in the phase diagram.

To optimize repairing we need to minimize this metric. Figure~5 shows
the solution to the minimization problem, and a detailed discussion is
provided in Methods. The different colors in Fig.~5 correspond to the
different optimal repair strategies, which depend on the failure state
of the system. If the system is initially at point A, both networks are
in a low activity state, i.e., they are non-functional. Our goal is to
decrease $p_A^*$ and $p_B^*$ and arrive to the region where the system
is fully recovered (the green region) by performing a minimal number of
repairs, i.e. minimal $N_{\rm rep}$. We find that for any point in
the red region there are actually two closest points in the green region,
at an equal Manhattan distance away from the red region point.
These two points are the triple points R1 and R2 shown in Fig.~5, which also correspond to the triple points in Fig.~2b.
Although R1 may be closer to point A than R2 by Euclidian distance, 
the Manhattan distance is the same. Thus two
equally good repairing strategies are available.  One involves
allocating more node repairs to network A, and the other allocating more
repairs to network B. For the yellow regions (points B and C), the
closest points by Manhattan distance are R1 (for point B) or R2 (for point
C). Here only one triple point represents the optimal solution. Note
that the path samples in Fig.~5 are ``zig-zag'' in shape (to highlight that
we are minimizing $|\Delta p_A^*|+|\Delta p_B^*|$), but even when a
diagonal path (direct straight line) to a triple point is used, the
Manhattan distance is the same.  For the dark blue regions (points D and
G), the optimal strategy is to decrease $p_B^*$ only, until the system is
recovered.  Similarly, for the light blue regions (points E and F), the
optimal strategy is to decrease only $p_A^*$.

From our optimal repairing strategy analysis we find that the order of
repair (the specific path taken between the initial point and final point) 
does not affect the final result. Minimizing the Manhattan distance only
determines the optimal destination point. Therefore, there is actually a set 
of paths corresponding to equally optimal repairing processes. 
%However, if we value early partial results during the repair process 
%(for example, if we appreciate to have one of the networks repaired 
%as quickly as possible), the definition of ``optimal'' may be further restricted,
%and it may be optimal to choose those paths from the set of optimal paths,
%that allow the quick recovery of subsystems, i.e., individual networks.

\subsection{States and transitions in Real World Networks}

In relatively small networks ($N\approx 10$--1000) fluctuations are very
large. Thus, in small network systems exhibiting multistability it is possible to
observe phase flipping \cite{Derek, flip, BP} between different states.
Figure~6a shows the fraction of active nodes for both networks, in time,
for a symmetric choice of parameters, $p_A^*=p_B^*=0.21$,
$r_A=r_B=0.60$, and $r_d=0.15$, when each network has only $N=100$
nodes. Large fluctuations cause the system to jump between the different
states allowed for this set of parameters.  Note that interdependent
links cause the two networks to have partially dependent and correlated
dynamics. Very often a transition in one network triggers a transition
in the other. In Figure~6a we can identify examples of all four global states: 22,
11, 21 and 12. For example, at time $t\approx 400$ both networks are in
the high activity state (11), while at $t\approx 620$ network A is in
the low activity and network B in the high activity state (21).  Because
a controlled experimental changing of such parameters as $p_A^*$ or $p_B^*$ is
usually impossible or hardly accessible in real-world networks, we can exploit the
phenomenon of phase flipping, use it as a probe to explore different
layers of the phase diagram, and verify the existence of well-defined
states and the transitions between them in a real-world network system.
By observing the dynamics in a selected real-world interacting network,
we find evidence of rapid transitions between different states (Fig.~6b)
that strongly resemble the spontaneous phase switching seen in our model
(Fig.~6a).

\begin{figure} [!htbp]
\centering \includegraphics[width=0.48\textwidth]{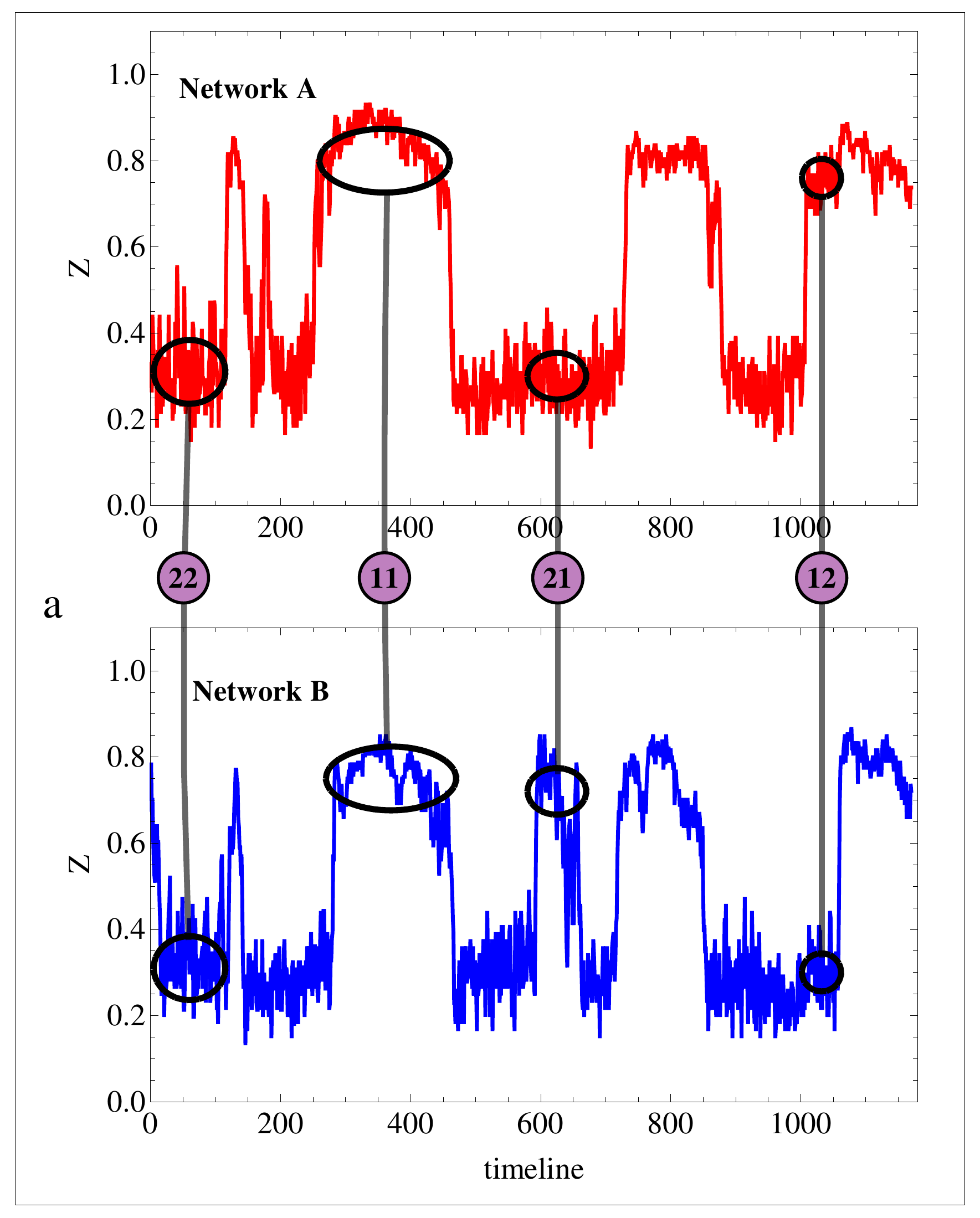}
\centering \includegraphics[width=0.48\textwidth]{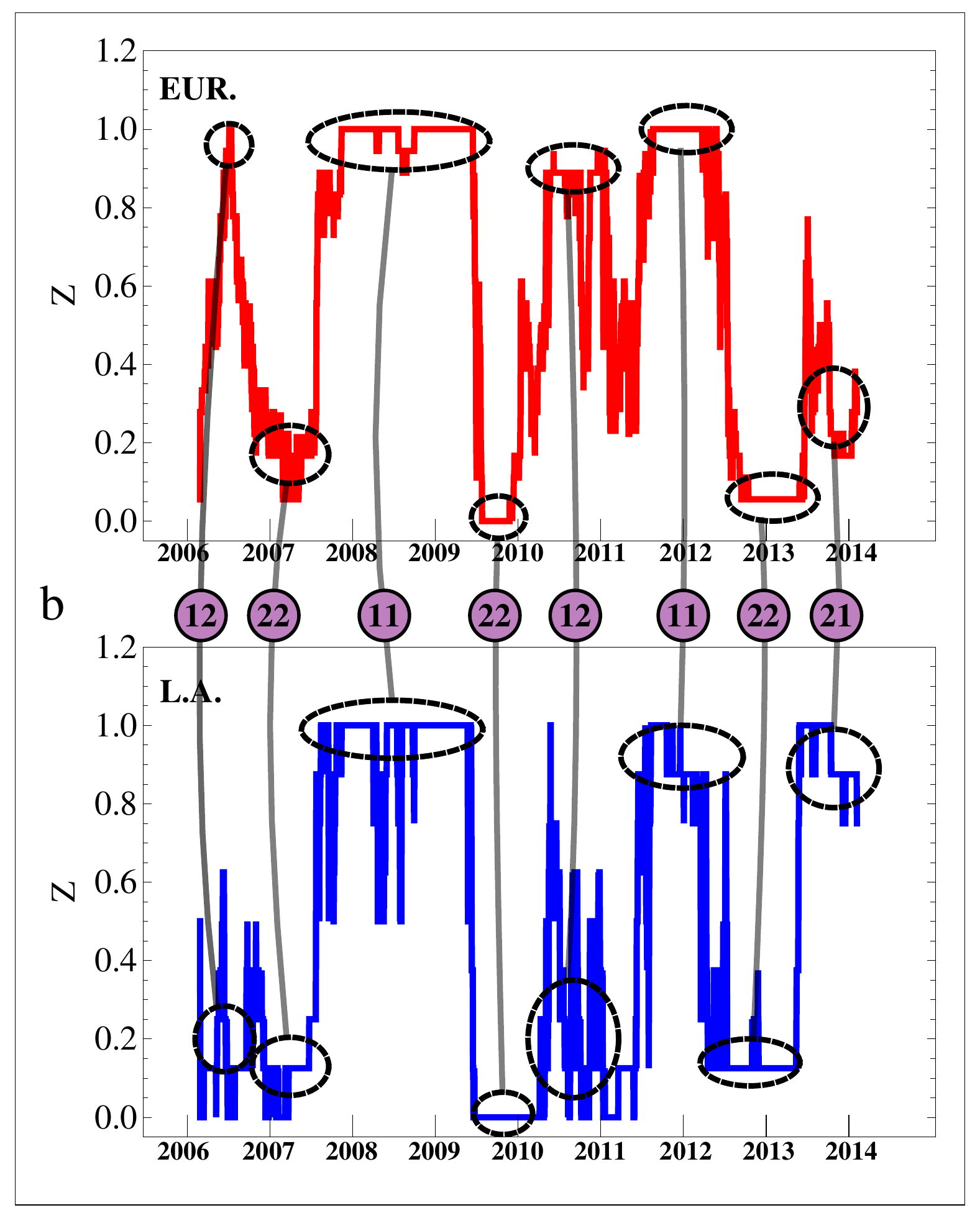}
\caption{ {\bf Collective states in simulated and real interacting
    networks}. {\bf a} Simulation of the networks' dynamics, activity
  versus time, for $N=100$ and failure parameters
  $p_A^*=p_B^*=0.21$, $r_A=r_B=0.60$, $r_d=0.15$, shows the switching of
  the system between four different states.  We can easily identify four
  collective states - 11, 22, 11, and 21.  {\bf b} Dynamics of two CDS
  geographical networks consisting of 18 European and 8 Latin American
  countries, shows very similar elements of the behavior: individual
  networks switching between well defined high activity and low activity
  states, as well as the correlated collective behavior of the two
  networks in interaction.  We identify collective states 11, 22, 12 and
  21 and mark them with connected black ovals.  Note that since the CDS
  value grows with risk, a higher activity in a CDS network corresponds
  to bad economic news.  }
   \label{6}
\end{figure}

To test our model with a real-world example we analyze the sovereign
debt 5-year credit default swaps (CDSs) of 26 different countries from
two geographical regions, Europe and Latin America. The full list of the
countries given in Methods, Part 5 represents all European and Latin
American countries that began to issue the CDS as early as 2005. A CDS
is a derivative contract that protects against the default risk of an
underlying reference asset as a result of a specific credit event, a
kind of insurance against credit default. In a CDS the buyer pays the
seller a premium for the recovery of a credit loss in case of
default. The higher the risk of default, the higher the premium, so the
value of a governmental CDS reflects the size and probability of a
potential loss for an investor in governmental bonds of a particular
country. A more detailed explanation of CDSs from an economic and
financial perspective is given in Methods, Part 5.  CDSs are leveraged
instruments (small changes in the underlying variables on which the
instrument is dependent can cause enormous changes in the value of the
instrument) and their values are very sensitive to both negative and
positive economic and political news emerging from various countries
- they reflect the sentiment or investor's perception of risk and fear 
about a particular country's economy. When one country is experiencing problems, 
this fear might affect the CDS values in other countries, 
usually in the countries of the same geographical region first,
and then in other countries. 
This behavior suggests that
we can treat countries as nodes and geographical regions (e.g., Europe
and South America) as interacting networks.

We examine the upward and downward movements of the CDS values in the 26
countries during the period June 2005--February 2014. We represent each
country with one node that can have two states: active or failed. Since
the CDS data is continuous, and in our model we have binary node states,
we perform the following mapping to produce a binary state for each
country. For each time $t$, we consider the interval [$t-252,t$] of 252
business days (this number is usually taken as the number of business
days in a year).  If the CDS of a country has a net increase during that
period, we consider the node of the country to be active at $t$. If it
does not, it is inactive.  Figure~6b shows the interaction of the two
geographical CDS networks: Latin America and Europe. First we note that
the networks indeed spend most of their time having either a very high activity 
or a very low activity (i.e., there are two well-defined
single-networks states).  We also observe that because of interactions
between the two networks they can share transition moments between high
and low activity, but sometimes these transitions occur
independently. This behavior is very similar to the model behavior observed in our simulations,
Fig.~6a. We conclude that our network model successfully captures the
behavior of this real network, and it represents a plausible model to
explain the most important elements of its evolution.

\section{METHODS}

{\bf 1. Damage conductivity parameters}. Parameters $r_A$ and $r_B$ are
introduced because they describe how easily the damage is spread 
through the network. When $r=0$ there is no damage spread between the
nodes, and when $r=1$ there is perfect damage conduction. Assuming that
external failures occur with certainty would mean fixing $r$ to be equal to 1. 
In the case of a single network with recovery it has been shown \cite{Derek} that many
important phenomena (e.g., spontaneous recovery) are lost when
$r=1$. The most interesting parts of the phase diagram are in fact where
$r$ is far from 1.

\bigskip  

{\bf 2. Mean field theory}. Fractions $a_A$ and $a_B$ denote the
fraction of nodes that are failed due to any of the three types of
failures: internal (I), external (E), or dependency failure (D).  We
denote the probabilities that a node at a time of observation
experiences a failure of I, E, or D type as $P$(I), $P$(E), and $P$(D),
respectively.  As a first approximation, we assume that these failures
are mutually independent events. Considering network A first, we write an expression
for the probability $a_{A,k}$ that a node of degree $k$ in network A has
failed.  The node can fail due to I, E, or D events or to a combination
of them. Using the inclusion-exclusion principle for independent events,
we write
\begin{equation}
a_{A,k}=P(I)+P(E)+P(D)-P(I)P(E)-P(I)P(D)-P(E)P(D)+P(I)P(E)P(D).
\label{e3}
\end{equation}

Next, we separately calculate $P$(I), $P$(E), and $P$(D).

{\bf Calculating P(I)}, the probability that a randomly chosen node is
internally failed at the time of observation. $P(I)$ is also the average
fraction of internally-failed nodes in a network, since internal
failures are independent events.  This is a Poisson process on
individual nodes \cite{Poisson, Derek}, and therefore $P(I)=e^{-p_A
  \tau}$. Since parameters $p_A$ and $\tau$ come in this expression as a
product, we can replace them with a single parameter, $p_A^*\equiv e^{-p_A
  \tau}$, which is bounded and also has the property $0\leq p_A^* \leq
1$, and so $P(I)= p_A^*$ for network A.

{\bf Calculating P(E)}, the probability that a randomly chosen node with
degree $k$ has externally failed.  Focusing once again on network A, without a loss
of generality, we let $F(k)$ be the probability that a node of degree
$k$ in network A is located in a critically damaged neighborhood (where
fewer than $m+1$ nodes are active). By definition, the time-averaged
fraction of failed nodes (for any reason) in network A is $0\leq
a_{A}\leq1$. In a mean-field approximation, this is also the average
probability that a randomly chosen node in that network has
failed. Using combinatorics, we obtain $F(k, a_{A}) =
\sum_{j=0}^{m}{{k\choose{j}} a_{A}^{k-j}{(1-a_{A})}^j}$~\cite{Derek}.
The probability that a node of degree $k$ in network A has externally
failed is then $P(E)=r_AF(k, a_{A})$. An analogous result is valid for
network B.

{\bf Calculating P(D)}, the probability that a node has failed due to
the failure of its dependent counterpart node in the other network. For
network A, this probability is equal to the product of parameter $r_d$
and the probability that a counterpart node in B has failed: $P(D)=r_d
a_B$. In network B by analogy this probability is equal to $r_d a_A$.

Writing Eq.~(\ref{e1}) for both networks and inserting the results for
P(I), P(E), and P(D) after summing over all k (and noting $a_A=
\sum_{k}f(k) a_{A,k}$ and $a_B= \sum_{k}f(k) a_{B,k}$), we get a
system of two coupled equations that describes the system of networks,
\begin{eqnarray}
a_A&=&p_A^*+r_d a_B(1-p_A^*) + \sum_{k}f(k) F(a_{A})[r_A-p_A^* r_A -r_A
  r_d a_B+p_A^*r_A r_d a_B]\label{e4}\\ a_B&=&p_B^*+r_d a_A(1 -p_B^*) +
\sum_{k}f(k) F(a_{B})[r_B-p_B^* r_B -r_B r_d a_A+p_B^*r_B r_d
  a_A]. \label{e5}
\end{eqnarray}

\bigskip

{\bf 3. "Forbidden" transitions.} Transition lines for $12\rightarrow21$
and $21\rightarrow12$ do not appear in our phase diagram, and it is quite
easy to understand why. Let’s assume that the transition line for
$12\rightarrow21$ does exist. To obtain that transition, the idea would
be to simultaneously increase $p_A^*$ and
decrease $p_B^*$ (i.e., increase the damage in one part of the system,
and decrease in another part). Suppose we are in phase 12 and
infinitesimally close to the supposed transition line.
Considering the local geometry of this line, we may be able to observe 
its angle with respect to the $p_A^*$ axis. If a transition occurs when
increasing $p_A^*$ and decreasing $p_B^*$, the tangent on the
supposed line would have an angle of $\theta \in
[0,\frac{\pi}{2}]$. From here it follows that by increasing $p_A^*$ only, while keeping $p_B^*$
constant, we would also make a transition (cross the transition line). The only other possibility would
be that we were moving {\it along\/} the transition line, but this is
easy to disprove because it would imply that the transition does not depend on
$p_A^*$.  If increasing $p_A^*$ only, causes a transition, the transition
must end in state 22, not 21. This is because if we only increase $p_A^*$, we increase
damage to both network A (directly) and network B (indirectly, through
the interdependent links).
 
\bigskip

{\bf 4. Geometry of the Manhattan distance minimization problem}. The
optimal strategies shown in different colors in Fig.~5 are derived from
the geometrical reasoning shown in Fig.~7.  Figure~7a shows a plot of a
series of curves consisting of points at identical Manhattan distances
from point A (equidistant curves). They produce a ``diamond'' shape, and the minimal
Manhattan distance between point A and the green region translates into
the task of ``fitting'' the diamond so that it just touches the green
region and its center is at A. The diamond in Fig.~7a touches the green
region at two points---triple points, which are the solution to the
minimisation problem. Figure~7b shows the solution for point F in the
light blue region. Here the solution suggests a different
strategy---decreasing only $p_A^*$.

\begin{figure} [!htbp]
\centering \includegraphics[width=0.475\textwidth]{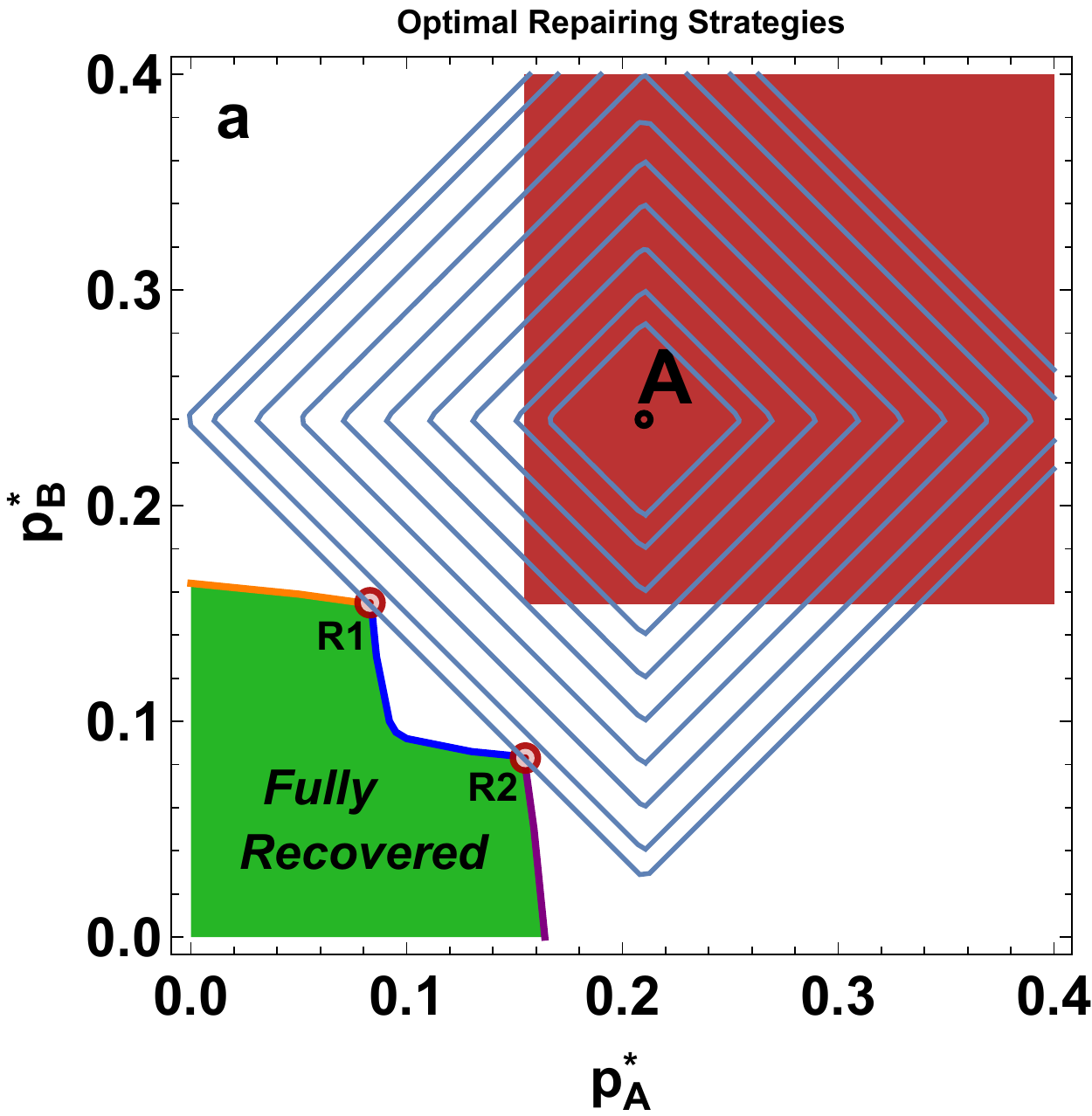}
\centering \includegraphics[width=0.49\textwidth]{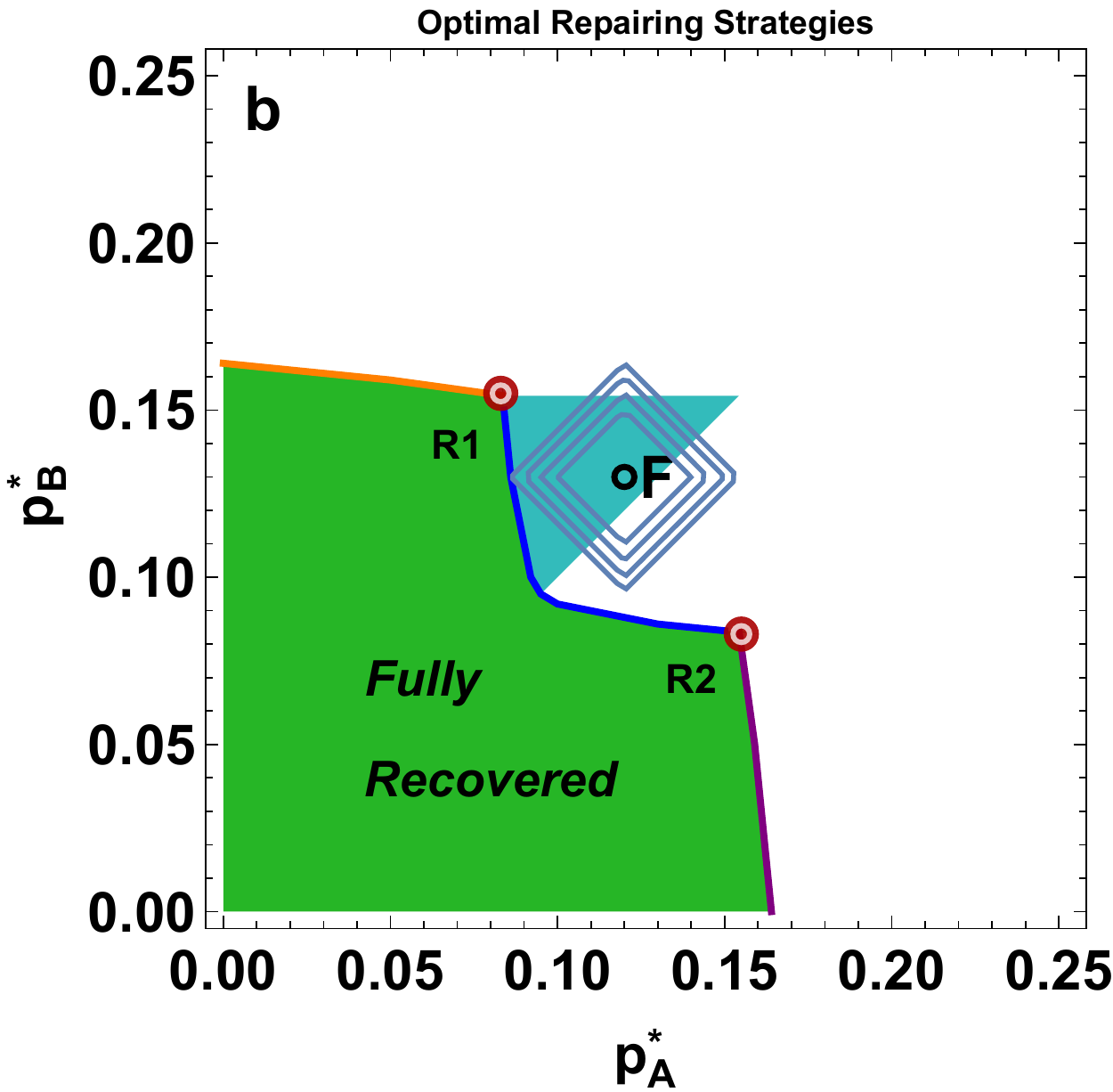}
\caption{ {\bf Minimum Manhattan distance problem}. 
{\bf a} For the red sector, fitting the largest ``diamond'' barely touching the green region
and having its center at point A, suggests there are two equally optimal 
solutions to the minimization problem. {\bf b} The same geometrical construction
for point F in the light blue region, suggests a unique solution: decreasing $p_A^*$.
   }
   \label{7}
\end{figure}

\bigskip

{\bf 5. Credit default swaps.}  Figure 6b shows an analysis of 5-year sovereign
debt CDSs for a set of European countries: France, Germany, Italy, Spain, Portugal, 
Belgium, Austria, Denmark, Sweden, Greece, Ukraine, Hungary, Poland, Croatia,
Slovenia, Romania, Bulgaria, and Slovakia. This is the set of European countries that had a
sovereign debt CDS in 2005. The set of Latin American countries we
analyze consists of Brazil, Colombia, Argentina, Mexico, Venezuela,
Chile, Peru, and Panama. A CDS is typically used to transfer the credit
exposure of fixed income products from one party to another.  The buyer
of the CDS is then obligated to make periodic payments to the seller of
the CDS until the swap contract matures. In return, the seller of the
CDS agrees to compensate (pay off) the seller who holds this third party
debt if this (third party) defaults on the issued debt.

A CDS is, in effect, an insurance against non-payment of a debt owed by
a third party. The buyer of a CDS does not have to hold the debt of the
third party but can speculate on the possibility that the third party
will indeed default, and the buyer can purchase the CDS for this
speculative purpose. CDSs were developed in the 1990s and, given their simple
structure and flexible conditions, they are now a major part of the credit
derivative activity in the OTC market used to hedge credit risk.
One of the most important aspects of a CDS is the definition of the
``credit event'' that triggers the CDS. These events include bankruptcy,
obligation acceleration, obligation default, failure to pay, repudiation
(moratorium), and restructuring.  In the case of the sovereign bond
market, the last three are typically included in the contracts.
CDSs are used by investors to hedge exposure to a fixed income instrument, 
to speculate on likelihood of a third party (reference asset) default, 
or to invest in foreign country credit without currency exposure.

\subsection*{Acknowledgments}

\noindent 
We thank the DTRA, NSF (grants CMMI 1125290, CHE-1213217 and “SES 1452061”), Keck Foundation, European Commission FET Open Project (“FOC” 255987, “FOC-INCO” 297149) and Office of Naval Research for financial support.
S.H. acknowledges the European LINC and MULTIPLEX (EU-FET project 317532) projects, the Deutsche Forschungsgemeinschaft (DFG), the Israel Science Foundation, ONR and DTRA for financial support. 
L.A.B thanks UNMdP and FONCyT, PICT 0429/13 for financial support.
S.L.C. gratefully acknowledges the financial support of the Fulbright Program for visiting scholars.
A.M. thanks Bijeli Zeko for useful discussions.

\end{document}